\documentclass[twocolumn]{article}
\usepackage{url}
\usepackage{amsmath}
\usepackage{multirow}
\usepackage{graphicx}
\usepackage{epstopdf}
\usepackage{lineno}
\usepackage{wrapfig}
\usepackage[affil-it]{authblk}
\usepackage{natbib}

\modulolinenumbers[5]

\newif\ifdraft
\draftfalse

\begin{document}


\newcommand{\Eqnref}[1]{Eq.~(\ref{#1})}
\newcommand{\Eqnsref}[2]{Eqs.~(\ref{#1}-\ref{#2})}
\newcommand{\EqnTworef}[2]{Eqs.~(\ref{#1}) and (\ref{#2})}
\newcommand{\Figref}[1]{Fig.~\ref{#1}}
\newcommand{\Figsref}[2]{Figs.~\ref{#1}-\ref{#2}}
\newcommand{\FigTworef}[2]{Figs.~\ref{#1} and \ref{#2}}
\newcommand{\Tabref}[1]{Table~\ref{#1}}

\providecommand{\etal}[0]    {et al.}
\newcommand{\eg}[0]      {e.g.}
\newcommand{\ie}[0]      {i.e.}
\newcommand{\etc}[0]     {etc.}

\newcommand{\entrop}      {{\mathbf H}}
\newcommand{\mutinfop}      {{\mathbf I}}
\newcommand{\miop}{\mutinfop}
\newcommand{\teop}        {{\mathbf T}}
\newcommand{\te}[2]      {\teop_{{#1}\to{#2}}}
\newcommand{\gteop}       {{\mathbf G}}
\newcommand{\bracr}[1]   {\left({#1}\right)}
\newcommand{\mutinf}[2]  {\mutinfop\!\bracr{{#1}:{#2}}}
\newcommand{\tlogtwo}{\log_2}
\newcommand{\gte}[2]      {\gteop_{{#1}\to{#2}}}
\newcommand\bv           {\boldsymbol{v}}
\newcommand\bx        {\boldsymbol x}
\providecommand{\e}[1]{\ensuremath{\times 10^{#1}}}

\newcommand{\todonew}[1]{\textcolor{red}{***#1***}}
\newcommand{\mycolrule}[1]{\\[-7.5pt]\cline{#1}\\[-7.5pt]}

\title{Review of Data Structures for Computationally Efficient Nearest-Neighbour Entropy Estimators for Large Systems with Periodic Boundary Conditions
}

\author{Joshua M. Brown \thanks{Email: \texttt{jbrown@csu.edu.au}; Corresponding author}} \affil{School of Computing \& Mathematics, Charles Sturt University, Panorama Avenue, Bathurst, New South Wales, 2795, Australia}

\author{Terry Bossomaier} \affil{Centre for Research in Complex Systems, Charles Sturt University, Panorama Avenue, Bathurst, New South Wales, 2795, Australia}

\author{Lionel Barnett} \affil{Sackler Centre for Consciousness Science, Department of Informatics, University of Sussex, Brighton, United Kingdom}

\maketitle

\begin{abstract}
Information theoretic quantities are extremely useful in discovering relationships between two or more data sets. One popular method---particularly for continuous systems---for estimating these quantities is the nearest neighbour estimators. When system sizes are very large or the systems have periodic boundary conditions issues with performance and correctness surface, however solutions are known for each problem. Here we show that these solutions are inappropriate in systems that simultaneously contain both features and discuss a lesser known alternative solution involving \emph{Vantage Point} trees that is capable of addressing both issues.

\end{abstract}

\section{Introduction}
\label{sec:intro}
Information Theory provides three useful metrics for determining relationships between data sets. \emph{Mutual~Information}~($\miop$)~\citep{Shannon48} measures the shared uncertainty between two variables and is often used as a marker for second-order phase transitions \citep{Matsuda96,harre09:epl,Wicks07}. Meanwhile, \emph{Transfer~Entropy}~($\teop$)~\citep{Schreiber00} and \emph{Global~Transfer~Entropy}~($\gteop$)~\citep{Barnett13} track the flow of information from one or more data sets onto another, with Barnett~\etal~demonstrating that $\gteop$ can be used as a predictor for an oncoming phase transition in the \citet{ising25} spin model when system temperature is reduced over time.

If the underlying distribution for the data is known, these quantities can be straightforward to calculate. However, in general this distribution is unknown and can only be approximated from sampled data, thus requiring \emph{Entropy Estimators}. Two well known approaches are the plug-in and nearest-neighbour (NN) estimators.

While simple, plug-in estimators tend to be less accurate, particularly for continuous variables~\citep{Ross14}~which are the focus of this paper. A high quality alternative is the Kozachenko-Leonenko entropy estimator~(\citeyear{Kozachenko87}) which uses nearest neighbour statistics of realisations of a data series. \citet{Kraskov04} extended this for estimating $\miop$, with $\teop$ and $\gteop$ estimators further derived by \citet{Gomez-Herrero15}. For higher dimensional data, NN estimators become even more appropriate as plug-in estimators experience a combinatorial explosion in storage requirements. 

On the other hand, the time complexity for the plug-in estimator scales linearly on the number of points, $N$, while NN estimators---using a direct method, where distances between each and every point are considered---scale on the order of $N^2$, which becomes significant for large $N$. This problem is widely encountered in the field of computer graphics and collision detection in computer games. The accepted solution for these scenarios are spatial partitioning data structures---such as well-known BSP Trees~\citep{Fuchs80} and KD Trees~\citep{Bentley75}---resulting in a \emph{``broad-phase''} collision detection step, which provides the ability to discard large swathes of points from consideration with relatively little computation.

KD Trees are directly applicable to the nearest neighbour searches required by the NN estimators, and are commonly applied when data sets are large. However, when the data lies in a space with periodic boundaries---\eg, headings of observed objects, where $2\pi=0$---the basic KD Tree fails. While there are modifications to overcome this, it is shown later that ultimately a more appropriate data structure is needed and available in the \emph{Vantage Point}~(VP) Trees developed---independently---by~\citet{Uhlmann91} and \citet{Yianilos93}.

This paper serves to highlight the relative computational performance of the VP Tree compared with a modified KD Tree for the specific use case of measuring $\miop$, $\teop$ and $\gteop$ of large periodic systems via a NN estimator. This use case also presents a unique constraint in that the search structures are typically constructed and used just one time---meaning construction costs are just as important as search costs.

To establish the accuracy of the VP Tree approach, tests are performed against periodic and aperiodic canonical distributions as done by~\citet{Kraskov04}. To further analyse the use of the VP Tree for large data sets with periodic boundary conditions, data sets---generated by the widely-used Vicsek model of self-propelled particles~\citep{Vicsek95}---are analysed. Lastly, work by \citet{Gao15} stresses that for accurate estimation, the NN estimator requires a number of points scaling exponentially with the true $\miop$. A decimation of the Vicsek data is employed to investigate this issue.

The three aforementioned information theoretic metrics are widely used in the scientific community, where mutual information has been used in many applications, including the study of phase transitions in financial markets~\citep{harre09:epl}, while the range of applications for transfer entropy is building steadily, from the study of physics of fundamental systems, such as the Ising spin model~\citep{Barnett13} to cellular automata~\citep{liz08a}, random Boolean networks\citep{liz12b}, and swarms~\citep{wang12a}. Ensuring the accuracy of the VP Tree method in periodic systems is important as it will allow processing of larger data sets with fewer computational resources---which is particularly important in the age of big data.

\section{Methods}
\label{sec:methods}

The most simple of the three quantities, $\miop$, measures the amount of uncertainty common to two variables, say the headings of particles in a simulation. The input for an estimator for $\miop$ is a set of data points, $(x,y)$, drawn from the two variables. $\teop$ instead measures the flow of information between two variables, that is, the reduction in uncertainty of $X$ given past knowledge of $X$ and $Y$. Estimators for $\teop$ thus requires a time series of data points $(x_t,x_{t-1},y_{t-1})$. Finally, $\gteop$ measures a similar statistic as $\teop$, but is used in the case where $X$ is influenced by multiple variables $Y^{d-2}$, and therefore input to these estimators is $d$-dimensional data set, $(x_t,x_{t-1},y^1_{t-1},\ldots,y^{d-2}_{t-1})$. 

The standard forms of the three metrics are~\citep{Shannon48,Schreiber00,Barnett13}:
\begin{equation}
\mutinf{X}{Y}=\entrop(X)+\entrop(Y)-\entrop(X,Y),
\label{eq:miEntropForm}
\end{equation}

\begin{equation}
\begin{split}
        \te{Y}{X} &=-\entrop(X_t,Y_{t-1},X_{t-1}) + \entrop(X_t,X_{t-1}) \\ 
        &\quad + \entrop(X_{t-1},Y_{t-1}) - \entrop(X_{t-1}),
\label{eq:teEntropForm}
\end{split}
\end{equation}

\begin{equation}
\begin{split}
        \gteop &=\frac1N \sum_i^N \te{Y_i'}{X_i}\,,
\label{eq:gteEntropForm}
\end{split}
\end{equation}
where $\entrop(Z)=-\displaystyle\sum p(z) \tlogtwo p(z)$. $\miop$ and $\teop$ are pairwise quantities, and as such, can be calculated using just two variables---and further averaged over every pairwise combination of variables if the data set contains many interacting variables. $\gteop$ on the other hand is a global quantity, so each random variable is considered in turn for the target variable. $Y_i'$ is thus the set of all influencing variables on the chosen target, $X_i$. 

Throughout this work, $X$ and $Y$ are drawn from a variety of sources where: (a) analytic results are available, (b) prior results for metrics are available, or (c) statistics are unknown and hard to estimate. A collection of canonical distributions with analytic results and distributions with prior results are provided in Tables~\ref{tab:ksgEstimateAndExact}~and~\ref{tab:ksgClosedForm}. Amongst these distributions is the ubiquitous Gaussian distribution, as well as the Von Mises distribution which is accepted as the circular analogue of the Gaussian distribution. The Vicsek Model (described in \S\ref{subsec:vicsekModel}) is used to generate difficult-to-estimate data.

There are some optimisations available for $\gteop$ depending on the nature of the data. If the variables of the data set are indistinguishable then $\gteop$ can be calculated as a single ensemble calculation treating it as a multivariate $\teop$ calculation, rather than $N$ $\teop$ calculations. This technique is employed in \S\ref{subsec:nnAccuracy}. A second such optimisation is dependent on how variables influence each other. In the case of the Vicsek model used later, variables influence one another in an aggregate fashion from which a consensus variable, $Y^c$, can be constructed. Thus $\gteop$ can be calculated as the 3-dimensional $\gte {Y^c}{X}$, rather than the above $d+2$-dimensional form. This reduction is utilised in \S\ref{subsec:performance}-\ref{subsec:vicsekChecks}.

\subsection{Nearest neighbour method}
\label{subsec:nnMethod}

The plug-in estimation approach uses these data sets to determine the underlying distribution often via a histogram. As mentioned in the introduction, once this distribution is revealed it is often trivial to solve the above equations. However, working with continuous data presents certain issues for traditional plug-in histogram entropy estimation. For example, tuning the bin width parameter for discretising the continuous data is susceptible to the number of points available, which can introduce error at high or low noise if the bin width is too low or high, respectively. While adaptive binning can mitigate this somewhat, it is not always an option, particularly when the data covers the entire domain.

\begin{figure}[!h]
\centering
  \includegraphics[width=\columnwidth]{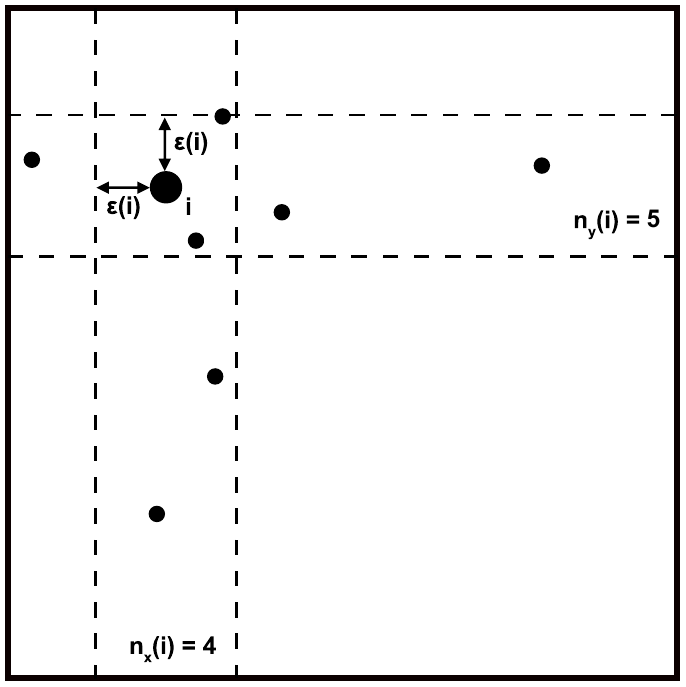}
  \caption{Determining the $k$th nearest neighbour---$k=2$ in this case---and $\epsilon(i)$ for some point $i$ and then count the number of points in the marginals strictly within these bounds, with $n_x(i)=4$ and $n_y(i)=5$, after~\citet{Kraskov04}.}
  \label{fig:nneIntro}
\end{figure}

Instead, one can use Nearest-Neighbour Entropy Estimators. These estimators are based on the (continuous) statistics of the $k$ nearest neighbours to each data point and the distribution of points along marginal dimensions as introduced by~\cite{Kozachenko87}. In these methods, the max-norm distance, $\epsilon(i)$, is found for each point $i$ and its $k$th nearest neighbour (kNN search). A fixed range (FR) search is then performed to count the number of points within the (hyper-)\emph{``ball''} defined by $\epsilon(i)$ from $i$ along the marginal spaces. These values are then used to compute the above information theoretic quantities. See \Figref{fig:nneIntro} for an example of collecting nearest neighbour information. 

\citet{Kraskov04} provide the nearest neighbour estimator for $\miop$ as:

\begin{equation}
\begin{split}
\mutinf{X}{Y} &= \psi(k) + \psi(N) \\&\quad - \langle\psi(n_x(i) + 1) + \psi(n_y(i) + 1)\rangle,
\label{eq:miKraskovI1}
\end{split}
\end{equation}
where $\psi(\cdot)$ is the digamma function, $n_z(i)$ is the count of points within the distance $\epsilon(i)$ from point $i$ in the marginal space $z$ (either $x$ or $y$), and $\langle\ldots\rangle$ is the average over all points.

The extension to this for $\teop$ and $\gteop$ is provided by \citet{Gomez-Herrero15}, and requires computing:
\begin{equation}
\begin{split}
\te Y X &= \psi(k) - \langle\psi(n_{xw}(i) + 1) + \psi(n_{xy}(i) + 1) \\&\quad- \psi(n_x(i) + 1)\rangle,
\label{eq:teGomez}
\end{split}
\end{equation}

\begin{equation}
\begin{split}
\gteop &= \psi(k) - \langle\psi(n_{xw}(i) + 1) + \psi(n_{xy'}(i) + 1) \\&\quad- \psi(n_x(i) + 1)\rangle,
\label{eq:gteGomez}
\end{split}
\end{equation}
where $x$ and $y$ represent the state at  $t-1$,  $w$ represents the current state of $x$ at $t$ and $y'$ represents the set of $Y^{d-2}_{t-1}$ variables used for the $\gteop$. $n_{xw}$ is thus the number of points that exist within the distance $\epsilon(i)$ from point $i$ when only considering the $x_{t-1}$ and $x_t$ coordinates of the data set.

Each of the above quantities requires $N$ kNN searches and either $2N$ or $3N$ FR searches. In a na\"ive approach, each of these searches is a $\mathcal{O}(N^2)$ operation, clearly making this approach more computationally expensive than simple plug-in estimators. However, by choosing more appropriate searching algorithms we can significantly reduce the cost of these estimators such that they become a feasible approach.

\subsubsection{Underlying data structure}

The na\"ive approach for finding the $k$th nearest neighbours and the marginal neighbour counts for point $i$ is to calculate the distance between $i$ and every other point, and only keep those points passing the criteria---\ie, amongst the $k$ closest points, or within distance $\epsilon(i)$. This approach requires $N$ comparisons for all $N$ points, resulting in a computational complexity of $\mathcal{O}(N^2)$ and as such is not a suitable approach when $N$ is large.

\begin{figure}[!h]
\centering
  \includegraphics[width=\columnwidth]{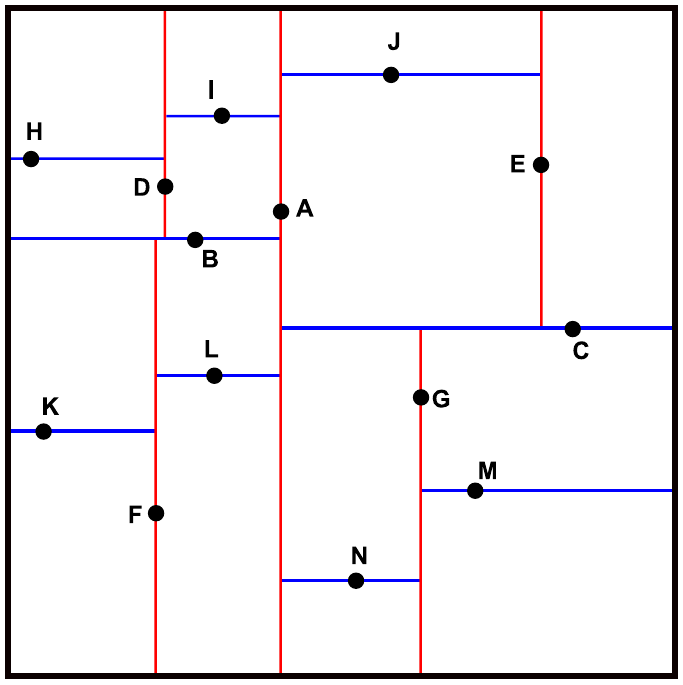}
  \caption{KD Tree constructed via lines intersecting each point segregate the local area of the point into so-called \emph{front} and \emph{back} areas. In this example, point A is the root node in the binary tree, with points B and C being its immediate child nodes, as seen in \Figref{fig:kdTreeTree}.}
  \label{fig:kdTreeExample}
\end{figure}

\begin{figure}[!h]
\centering
  \includegraphics[width=\columnwidth]{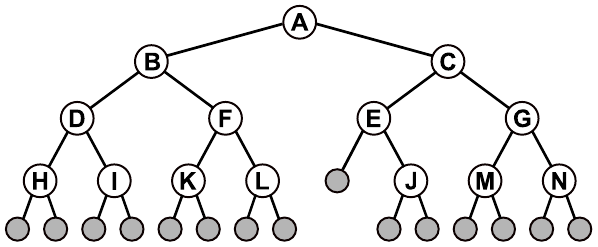}
  \caption{Tree structure created by subdivision of space in KD Tree. All nodes descending from the left of A appear in front of A in the space, while those descending from the right of A appear behind A. Grey leaf nodes represent the areas enclosed by the subdivisions. An algorithm for traversing this tree, presented by \citet{Friedman77}, allows finding the $k$th nearest neighbour in $\tlogtwo N$ time, rather than $N^2$.}
  \label{fig:kdTreeTree}
\end{figure}

A common solution to this problem in computer graphics are data structures such as the KD Tree~\citep{Bentley75}, which subdivide a space using lines to define front/back areas, as seen in \Figref{fig:kdTreeExample}. When considering point $i$, one can quickly cull large sections of space---those containing points which cannot possibly be among the $k$ nearest neighbours, or within the bounds $\epsilon(i)$---from further consideration. This approach reduces the complexity to $\mathcal{O}(N\tlogtwo N)$~\citep{Friedman77}. The subdivision process creates a tree hierarchy which can be seen in \Figref{fig:kdTreeTree}.

However, when the data lies in a space with periodic boundary conditions a structure such as a KD Tree will incorrectly cull points. Since the space wraps, a point is both in front of and behind any other point, while the KD Tree only considers points up to the boundary and not beyond. Three potential methods for correcting this---using Images, a hybrid involving na\"ive search and KD Trees, and VP Trees---are discussed below.

The images method duplicates and shifts the data---thus creating cloned images---such that when the KD Tree is constructed from the combination of the original data set and all images, it has the illusion of periodicity. This requires $3^d-1$ duplications for a $d$-dimensional space which can quickly become intractable, especially for $\gteop$. Note however, that typically only the \emph{shortest} distance between points $i$ and $j$ are considered and as such this can be reduced to $2^d-1$ by only duplicating the closest region (half/quarter/eighth/\etc) as needed. Some specific cases---perhaps where direction is a factor---might require $3^d-1$ full duplications, however, this will not be considered here, due to both the difficulty in imagining such a scenario---implying the data is both  circular and \emph{pairwise} ordered---as well as the fact that the two following approaches are only capable of solving the shortest distance form.

The second method---a hybrid of duplication and na\"ive search---requires only a single duplication, but can degenerate to a na\"ive search for select points. The image in this method---queried separately---is shifted such that points originally at the corners are now in the centre of the space, and vice versa---\ie, each dimension is shifted by half of its width. When querying $x_i$ (kNN or FR), if it is closer to the boundary than $\epsilon(i)$---\ie, there are possibly points just on the other side of the boundary closer than $\epsilon(i)$---the query is repeated in the image. The boundary check is repeated on the image result, and if it fails again, the query steps down to a na\"ive search. This method requires less duplication than the previous method but requires more computation per point.

\begin{figure}[!h]
\centering
  \includegraphics[width=\columnwidth]{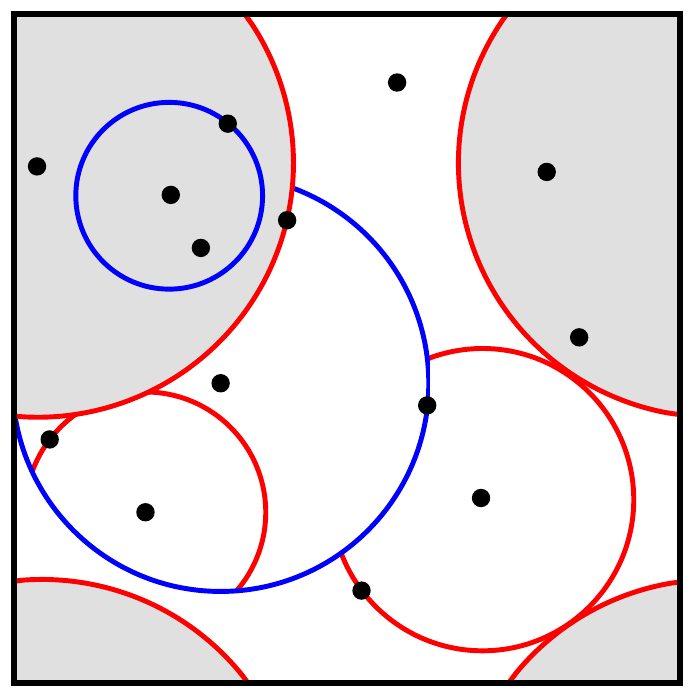}
  \caption{Circles centred on vantage points delineating near/far subsets. Note that child circles do not cross the boundaries of parent circles, as vantage points only consider points in the same subset as itself. Shaded region represents the near subset of the root node (top left), accounting for periodic boundary conditions.}
  \label{fig:vpTreeExample}
\end{figure}

Instead of wrangling the KD Tree to handle periodicity, the third approach is to use a different structure. As already discussed, KD Trees cannot deal with periodic conditions because the nature of front and back areas in periodic spaces is ill-defined. VP Trees work instead by choosing a point, $v$ (the vantage point), and then subdividing the space into those points nearer to $v$ and those points farther from $v$ than a given threshold. Since distances---and thus near/far points---are trivial to calculate in a periodic space, this structure can be used to attain efficient searching without resorting to any tricks. To create a balanced tree, the threshold is chosen such that it divides the points in half. The subdivision process is repeated on each new subset with new vantage points and thresholds chosen from only those points in the subregion. \Figref{fig:vpTreeExample} shows how this subdivision strategy works in an example space. Note that the hierarchical structure of this tree is the same as the KD Tree, just with the left children being near---instead of in front---and the right children being far.

One drawback to this near/far dichotomy is that a new tree is needed for each set of marginals as the dimension reduction can change the distance between points. That is, if the max-norm distance between $d$-dimensional realisations, $||x_i - x_j||_\infty$, is the distance along dimension $d_\alpha$, then a marginal subset that does not include $d_\alpha$ will have a different distance between $x_i$ and $x_j$. Similar arguments apply to other norms.

Thus a separate tree is needed for each kNN and FR search---totalling three for $\miop$ and four for $\teop$ and $\gteop$. However, the FR searches are independent of each other, thus if the kNN search is performed and the $\epsilon$-ball sizes stored the remaining trees can be used in serial. Since each tree is only used once (with all $N$ points processed), only one tree is ever needed at any given time, reducing the space requirements for this approach to $2N$, rather than $3N$ or $4N$.

When performing an FR search in just one dimension (\ie, for counting $n_x$ and $n_y$), it was found that performing a binary search to find the range of elements covered by $\epsilon(i)$---and thus the neighbour count---was faster than VP Trees. As such in the one dimensional searches, all test cases revert to this simpler algorithm. This also reduces the number of VP Trees required to one for $\miop$ and three for $\teop$ and $\gteop$.

The KD Tree implementation used in this paper was provided by the \texttt{ANN} library~\citep{Mount10}, which is a standard package in the area of nearest neighbour searches with over 330 citations on \emph{Google Scholar}, while the VP Tree implementation is a slightly modified version of the public domain code by \citet{Hanov12}. Our software is provided online. Both implementations are written in C++.

It should also be noted that other structures and approaches exist for solving this issue. For example, the periodic KD Tree library~\citep{Varilly14}~(written in Python) instead duplicates the query points rather than the data points. While this halves the storage requirements, each query point needs to be processed $2^d$ times (in the typical case) or $3^d-1$ times (in the general, if not odd, case), rather than 1 to 3 times as in the hybrid model. This can be an obstacle in the case of measuring information theoretic quantities where each data point is also a query point (\ie, $Q=N$ rather than $Q\ll N$). Another two  structures capable of addressing the kNN problem are Locality Sensitive Hashing (LSH)~\citep{Zhang13} and higher order Voronoi Diagrams (specifically, k-order)~\citep{Lee82,Dwyer91}. For the purposes of estimating information theory quantities however, these algorithms will not perform significantly better than the KD Tree or VP Tree due to space requirements, the one-shot nature of the estimation (that is, only using each structure once), and the distribution of data points. As such, the authors shall simply note their existence (and better performance in other use cases) and consider just the KD Tree and VP Tree structures for the remainder of this paper.

\subsection{Vicsek Model}
\label{subsec:vicsekModel}
The standard Vicsek Model (SVM)~\citep{Vicsek95} is a simple model that is widely used in collective motion studies. The simplicity of the model allows it to easily scale to large system sizes (both in terms of time and space). As such, it easily generates a large number of points---with periodic boundary conditions---perfect for testing VP trees. In the model, a flock of $M$ particles move in a continuous space (off-lattice) with periodic boundary conditions of linear size $L$. Particle density is defined as $\rho=\frac{N}{L^2}$. Particles move with constant speed, $s$, and are simulated for $\tau$ discrete time steps. Formally, the particles are updated according to:

\begin{align}
  \bx_i(t + \Delta t) &= \bx_i(t) + \bv_i(t)\Delta t \label{eqn:OVAposUpdate}\,, \\
  \theta_i(t+\Delta t) &= \langle\theta_i(t)\rangle + \omega_i(t) \label{eqn:OVAvelUpdate}\,,
\end{align}
where $\bx_i(t)$ is the position of particle $i$ at time $t$, $\bv_i(t)$ is its velocity, which is constructed with speed $s$ and heading $\theta_i(t)$. $\langle\theta_i(t)\rangle$ defines the average angle of all particles within $r=1$ units of $i$ (including itself) and $\omega_i(t)$ is a realisation of uniform noise on the range $[-\frac{\eta}{2},\frac\eta 2]$, where $\eta$ is our temperature variable and $0\leq\eta\leq2\pi$.

Particle alignment is measured via the instantaneous order (or disorder) using the parameter:

\begin{equation}
\varphi(t) = \frac1M \left| \sum^M_i\bv_i(t)\right|\,,
\end{equation}
where $0\leq\varphi(t)\leq1$. When $\varphi(t)=1$, the flock is completely aligned and collective motion has been attained, while $\varphi(t)=0$ represents the disordered, chaotic state of particles moving with no alignment. $\varphi(t)=1$ is associated with no noise in the system---\ie, $\eta\to0$---while $\varphi(t)=0$ occurs at high noise, $\eta\to2\pi$.

\begin{figure}[!h]
\centering
  \includegraphics[width=\columnwidth]{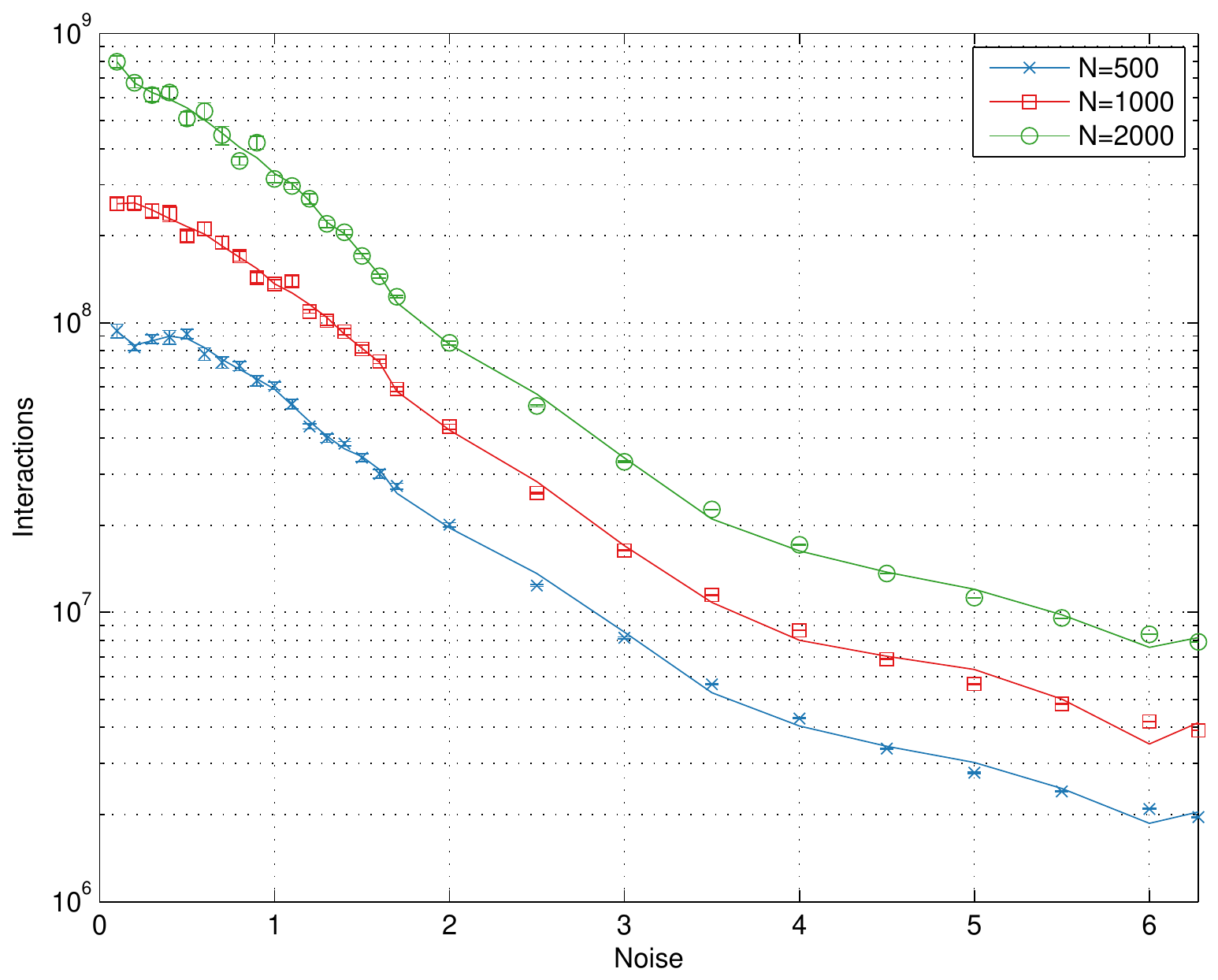}
  \caption{Logarithmic scale of number of interactions, $N_I$, for Vicsek system with parameters $\tau=5000, \rho=0.25, s=0.1$ for $N=500,1000,2000$. Error bars represent standard deviation over $10$ repetitions.}
  \label{fig:numInteractions}
\end{figure}

Note that while particle positions are indeed in a wrapped space, the random variables evaluated below are actually the heading \emph{angles} of neighbouring (interacting) particles. That is, the quantities describe the various reductions in uncertainty of a particle's heading given knowledge of its past heading and its neighbours' headings, not their positions. To measure the information theoretic quantities---using the notation from \Eqnsref{eq:miKraskovI1}{eq:gteGomez}---$X$ is defined as the heading of particle $i$, $Y$ as the heading of an interacting neighbour particle $j$, and $W$ as the next heading of particle $i$. In the case of $\miop$ and $\teop$, which are pairwise quantities, this is repeated for each interacting neighbour $j$ for every time step $t$. On the other hand, $\gteop$ measures all $j$ interacting neighbours in one multi-dimensional variable $Y'$, and thus generates just a single point. Particle indistinguishability is employed and repeated the above for all $i$, coalescing all points into a single point cloud. The pairwise quantities generate a separate data point for all $(i,j)$ interacting pairs, which is not constant over time or noise. See \Figref{fig:numInteractions} for a visualisation of how the number of interacting pairs changes as a function of noise temperature, $\eta$. $\gteop$ on the other hand generates a single point per $i$ for all of its interacting neighbours at time $t$ and thus contains precisely $\tau M$ data points at all noise values.

There is one major difficulty in calculating $\gteop$ however. $\gteop$ uses all interacting neighbours, $N_I$, to particle $i$ at time step $t$ to define the multivariate $Y'$---\ie, $d=N_I$. $N_I$ is not constant which causes issues when collecting all realisations into a single point cloud for calculation. One solution is to pick an arbitrary value of $d$---say the minimum, maximum or mean of $N_I$---however these options will miss pertinent variables and thus give an inaccurate result, or will include irrelevant particles and thus increase the complexity of the problem. The alternative is to note that particles only influence each other via the consensus heading, $\langle\theta_i(t)\rangle$, and to use it instead of each particle individually. The latter approach is used for all Vicsek results below.

Finally, to provide an example for the earlier claim regarding adaptive binning techniques for plug-in estimators, note that the average heading of Vicsek flocks will proceed on a random walk about the unit circle over time. As such, the angles generated by each particle will be roughly uniform over each $2\pi$ marginal (although in the joint space, accumulates around the $x\approx y\approx w$ diagonal), degenerating adaptive bins to simple fixed-width bins.

\section{Results}
\label{sec:results}

\subsection{Accuracy of NN estimator}
\label{subsec:nnAccuracy}

\begin{table*}[hptb]
\centering
\caption{Estimated and exact (italicised) values to 4 decimal places for $\miop$, $\teop$, and $\gteop$ using $1\e4$ realisations for variables $X$, $Y$, $W$ drawn from the listed bi- and multi-variate distributions and auto-correlated processes. Evaluation performed over $500$ repetitions to establish the mean estimate---and thus bias---and standard error of the estimator (\dag~Multiplied by $1\e4$ for readability). Bias is dependent on number of realisations and covariance of random variables, with improved bias as realisations increase and covariance decreases. *Gaussian, LogNormal and Cauchy distributions performed with covariance, $r=0$ (\ie, $\Sigma=I$) and $r=0.9$. Parameters for Von Mises and Hahs \& Pethel evaluations found in text.}

\begin{tabular} {@{}cccccccc@{}} \mycolrule{1-8}
& & \multicolumn{6}{c}{Metric}\\
\mycolrule{3-8}
\multicolumn{2}{c}{Distribution*} & $\miop$ & $\text{s.e.}^\dag$ & $\teop$ & $\text{s.e.}^\dag$ & $\gteop$ & $\text{s.e.}^\dag$ \\ \mycolrule{1-8}

\multicolumn{1}{c}{\multirow{4}{*}{Gaussian}} &
\multicolumn{1}{c}{\multirow{2}{*}{$r=0.0$}} & $-1\e{-5}$ & \multirow{2}{*}{3.4} & $0.0001$ & \multirow{2}{*}{3.5} & $-0.0004$ & \multirow{2}{*}{3.2} \\ 
\multicolumn{1}{c} {} &
\multicolumn{1}{c}{} & $\emph{0.0}$ && $\emph{0.0}$ && $\emph{0.0}$ & \\ [3pt]
\multicolumn{1}{c} {} &
\multicolumn{1}{c}{\multirow{2}{*}{$r=0.9$}} & $0.8336$ & \multirow{2}{*}{5.1} & $0.1276$ & \multirow{2}{*}{3.9} & $0.1827$ & \multirow{2}{*}{4.0} \\
\multicolumn{1}{c} {} &
\multicolumn{1}{c}{} & $\emph{0.8304}$ & &$\emph{0.1270}$ && $\emph{0.1816}$& \\ [3pt]

\multicolumn{1}{c}{\multirow{4}{*}{LogNormal}} &
\multicolumn{1}{c}{\multirow{2}{*}{$r=0.0$}} & $5\e{-6}$ & \multirow{2}{*}{3.4} & $4\e{-6}$ & \multirow{2}{*}{3.4} & $-0.0003$ & \multirow{2}{*}{3.0} \\
\multicolumn{1}{c} {} &
\multicolumn{1}{c}{} & $\emph{0.0}$ && $\emph{0.0}$ && $\emph{0.0}$ &\\ [3pt]
\multicolumn{1}{c} {} &
\multicolumn{1}{c}{\multirow{2}{*}{$r=0.9$}} & $0.9773$ & \multirow{2}{*}{5.1} & $0.1326$ & \multirow{2}{*}{3.7} & $0.1906$ & \multirow{2}{*}{4.0} \\
\multicolumn{1}{c} {} &
\multicolumn{1}{c}{} & $\emph{0.9741}$ && $\emph{0.1314}$ && $\emph{0.1871}$ &\\ [3pt]

\multicolumn{1}{c}{\multirow{4}{*}{Cauchy}} &
\multicolumn{1}{c}{\multirow{2}{*}{$r=0.0$}} & $0.1933$ & \multirow{2}{*}{4.6} & $0.0843$ & \multirow{2}{*}{4.1} & $0.1121$ & \multirow{2}{*}{4.0} \\
\multicolumn{1}{c} {} &
\multicolumn{1}{c}{} & $\emph{0.2242}$ && $\emph{0.0827}$ && $\emph{0.1251}$ &\\ [3pt]
\multicolumn{1}{c} {} &
\multicolumn{1}{c}{\multirow{2}{*}{$r=0.9$}} & $1.0438$ & \multirow{2}{*}{7.2} & $0.1880$ & \multirow{2}{*}{4.6} & $0.2602$ & \multirow{2}{*}{4.9} \\
\multicolumn{1}{c} {} &
\multicolumn{1}{c}{} & $\emph{1.0545}$ && $\emph{0.2098}$ && $\emph{0.3067}$ &\\ [3pt]

\multicolumn{2}{c}{\multirow{2}{*}{Uniform}} & $-0.0005$ & \multirow{2}{*}{3.4} & $-9\e{-5}$ & \multirow{2}{*}{3.4} & $-0.0003$ & \multirow{2}{*}{3.2} \\
\multicolumn{2}{c}{} & $\emph{0.0}$ && $\emph{0.0}$ && $\emph{0.0}$ &\\ [3pt]

\multicolumn{2}{c}{\multirow{2}{*}{Uniform Wrapped}} & $-0.0005$ & \multirow{2}{*}{3.4} & $1\e{-5}$ & \multirow{2}{*}{3.4} & $-0.0002$ & \multirow{2}{*}{3.2} \\
\multicolumn{2}{c}{} & $\emph{0.0}$ && $\emph{0.0}$ && $\emph{0.0}$ &\\ [3pt]

\multicolumn{2}{c}{\multirow{2}{*}{Hahs \& Pethel}} & \multirow{2}{*}{N/A} & \multirow{2}{*}{---} & $0.1659$ & \multirow{2}{*}{3.7} & \multirow{2}{*}{N/A} & \multirow{2}{*}{---}  \\
\multicolumn{2}{c}{} && & \emph{0.1651} & \\ [3pt]

\multicolumn{1}{c}{\multirow{6}{*}{Von Mises}} &
\multicolumn{1}{c}{\multirow{2}{*}{a}} & 0.3489 & \multirow{2}{*}{4.3} & \multirow{2}{*}{N/A} & \multirow{2}{*}{---} & \multirow{2}{*}{N/A} & \multirow{2}{*}{---}  \\
\multicolumn{1}{c} {} &
\multicolumn{1}{c}{} & $\emph{0.3488}$ &&& & \\ [3pt]
\multicolumn{1}{c} {} &
\multicolumn{1}{c}{\multirow{2}{*}{b}} & 0.4384 & \multirow{2}{*}{4.5} & \multirow{2}{*}{N/A} & \multirow{2}{*}{---} & \multirow{2}{*}{N/A} & \multirow{2}{*}{---} \\
\multicolumn{1}{c} {} &
\multicolumn{1}{c}{} & $\emph{0.4384}$ &&& & \\ [3pt]
\multicolumn{1}{c} {} &
\multicolumn{1}{c}{\multirow{2}{*}{c}} &0.1940 & \multirow{2}{*}{4.3} & \multirow{2}{*}{N/A} & \multirow{2}{*}{---} & \multirow{2}{*}{N/A} & \multirow{2}{*}{---} \\
\multicolumn{1}{c} {} &
\multicolumn{1}{c}{} & $\emph{0.1928}$ &&& & \\
\mycolrule{1-8}
\end{tabular}
\label{tab:ksgEstimateAndExact}
\end{table*}

To ensure the integrity of the VP Tree implementation of the NN estimator, $\miop$, $\teop$ and $\gteop$ are estimated from data series sampled from distributions with known closed form entropies. \citet{Kraskov04} performed this test for $\miop$ against Gaussian data with multiple covariances, which is repeated and extended here. This establishes the accuracy of the estimator for the $\teop$ and $\gteop$ as well. The estimator is also tested against the Log-Normal, Cauchy (Via a Student's t distribution with $\nu=1$ degree of freedom) and Uniform distributions. For these tests---and all tests below---$k$ is set to $3$ as per the suggestion of \citet{Kraskov04}, who propose the choice of $k=2\text{--}4$ as it provides a good balance between statistical and systematic errors in the NN estimator. For larger systems, larger $k$ is allowed as systematic errors tend to 0. Simulations with $k=30$ confirmed no change in estimate, and so only $k=3$ is presented below.

To extend the test to account for periodic cases, all three metrics are tested against a wrapped uniform distribution and $\miop$ against the von Mises distribution (sine variant). Exact $\miop$ for the von Mises distribution is provided in Table 1 of \citet{Hnizdo08}, for three parameter sets: a) $\kappa_1=10,\kappa_2=15,\lambda=10$, b) $\kappa_1=15,\kappa_2=12,\lambda=12$, and c) $\kappa_1=12,\kappa_2=14,\lambda=-8$, with $\mu_1=\mu_2=\pi$ for all three sets.

Lastly, the estimator is tested against the autoregressive process described by \citet{Hahs13} (Example 1) which includes an analytical expression for $\teop$. Variances of the Gaussian noise terms in each process were $Q=1, R=2$ for $X$ and $Y$, respectively, with $a=0.9$ and $h_c=1$, and an initial $x$ value of $x_0=0$.

The estimated and exact values for these distributions can be seen in \Tabref{tab:ksgEstimateAndExact}. Estimates were measured from $1\e4$ realisations and repeated $500$ times. Note that as the number of realisations increases, the estimate values converge to the exact values, as observed in \citet{Kraskov04}. Closed form expressions for the information theoretic quantities can seen in \Tabref{tab:ksgClosedForm}. 

\begin{table*}[phtb]
\centering
\caption{Closed form analytic expressions for $\miop$, $\teop$, and $\gteop$. Note that closed form expressions for the Von Mises distribution and the Hahs \& Pethel auto-regressive process require more context than can reasonably be provided here and as such the reader is referred to the original material for this context.}

\begin{tabular} {lr} \mycolrule{1-2}

\multicolumn{1}{c}{Distribution} & \multicolumn{1}{c}{Closed Form}\\
\mycolrule{1-2}

$\miop_{Gauss}, \miop_{LogNormal}$ & $-\frac{1}{2}\log(1-r^2)$ \\
$\teop_{Gauss}, \teop_{LogNormal}$ & $\frac{1}{2}\Big[-\log|\Sigma_{xyw}| + \log|\Sigma_{xy}| + \log|\Sigma_{xw}|\Big]$ \\
$\gteop_{Gauss}, \gteop_{LogNormal}$ & $\frac{1}{2}\Big[-\log|\Sigma_{xy'w}| + \log|\Sigma_{xy'}| + \log|\Sigma_{xw}|\Big]$ \\
$\miop_{Cauchy}$ & $\log (8\pi) - 3 - \frac12\log r$ \\
$\teop_{Cauchy}$ & $4 - \log(16\pi) + \frac12\Big[-\log|\Sigma_{xyw}| + \log|\Sigma_{xy}| + \log|\Sigma_{xw}|\Big]$ \\
$\gteop_{Cauchy}$ & $3+\log\frac{\Gamma(\frac{1+d}{2})}{4\Gamma(\frac{d}{2})\sqrt{\pi}} - \frac{d+1}{2}\psi(\frac{d+1}{2}) + \frac{d}2\psi(\frac{d}2) + \frac12\psi(1)$\\&$+\frac12\Big[-\log|\Sigma_{xyw}| + \log|\Sigma_{xy}| + \log|\Sigma_{xw}|\Big]$ \\
Uniform & $0$ \\
Uniform Wrapped & $0$ \\
Hahs \& Pethel & See \citet{Hahs13} \\
Von Mises & See \citet{Hnizdo08} \\
\mycolrule{1-2}
\end{tabular}
\begin{flushleft}Where $r$ is the scalar covariance between $X$ and $Y$ and $\Sigma$ is the covariance matrix between the subscripted variables. Additionally, $\Gamma(\cdot)$ is the Gamma function while $\psi(\cdot)$ is the digamma function. Information sharing and flow are non-existent for uniform distributions by definition and thus become 0.
\end{flushleft}
\label{tab:ksgClosedForm}

\end{table*}

Given the results in \Tabref{tab:ksgEstimateAndExact}, and the converging nature as the number of realisations is increased, it is determined that the VP Tree implementation of the NN estimator is accurate.

\subsection{Comparison of underlying data structure}
\label{subsec:performance}

\citet{Yianilos93} and \citet{Friedman77} provide \emph{Big $\mathcal{O}$} analysis of VP Trees and KD Trees, respectively, showing both can be constructed and searched in $\mathcal{O}(N \tlogtwo N)$ time and use $\mathcal{O}(N)$ space. Here, estimation of the complexity of the hybrid KD Tree approach is provided, followed by empirical data comparing computation and space of the VP Tree and hybrid approaches.

As mentioned above, the metrics for the Vicsek model are calculated from a two or three dimensional space with periodic boundary conditions, with sides $S=2\pi$. To perform kNN searches the hybrid method constructs two KD Trees: $\kappa$ partitions the points $N_I$, while $\kappa'$ partitions the points $N_I'$---the original points shifted by $\pi$ along every dimension, with wrapping.

To refresh, the hybrid search works as follows: search the first tree, $\kappa$, for the $k$th nearest neighbour to $i$, which gives a distance, $\epsilon(i)$. If $i$ is closer than $\epsilon(i)$ units to the boundary then the search progresses to the equivalent point $i'$ in the second tree, $\kappa'$, to find a corresponding distance, $\epsilon'(i)$. If $i'$ is also closer than $\epsilon'(i)$ units to the boundary, the search reverts to a na\"ive linear search over all $N$ points.

Thus $4$ areas can be defined. First is the total area, $A_T=(2\pi)^D$, containing all points. Second is the inner area, $A_I$ given by the box of sides $2\pi-2\epsilon$ centred in the space, containing all points successfully processed by $\kappa$ in the first step. Following this is the border area, $A_B=A_T-A_I$, covering those points that $\kappa'$ will process (successfully or otherwise). Finally, the na\"ive area, $A_N$, contains those points that will ultimately be processed using the linear search, and is defined as the area within a max-norm distance of $\epsilon$ from the boundary midpoints, such that $A_N=D2^D\epsilon^2$.

Due to the tiered approach of the hybrid method, the time complexity will follow the form

\begin{equation}
N\tlogtwo N + \alpha N\tlogtwo N + \beta N^2 \,,
\end{equation}
where $\alpha$ is the proportion of points in the border area and $\beta$ is the proportion of points in the na\"ive area. Note that $0 \leq \beta \leq \alpha \leq 1$.

In the high noise Vicsek case, where $N_I$ is uniformly distributed with density $\rho_I=N/(2\pi)^D$ the average distance to the $k$th nearest neighbour is~\citep{Bhattacharyya08,Gaboune93,Hertz09}:

\begin{equation}
\begin{split}
\epsilon &= c\left(\frac{k}{\rho_I}\right)^{\frac{1}{D}} \\
&=c\left(\frac{k(2\pi)^D}{N}\right)^{\frac{1}{D}}\,,
\label{eq:avgDist}
\end{split}
\end{equation}
where $c=\frac{7}{15}$. This allows further expansion of $A_I, A_B$ and $A_N$.

For the high noise uniform case the proportions are simply $\alpha = A_B/A_T$ and $\beta = A_N/A_T$, which gives time complexity of

\begin{equation}
(N+4c\sqrt{kN}-4c^2k)\tlogtwo N + 8ckN
\label{eq:miComplexity}
\end{equation}
for the two-dimensional $\miop$ case and 
\begin{equation}
\begin{split}
&(N+6ck^{1/3}N^{2/3}-12c^2k^{2/3}N^{1/3}+8c^3k)\tlogtwo N \\&+ 24c^3kN
\label{eq:teComplexity}
\end{split}
\end{equation}
for the three-dimensional $\teop$ case.

In other distributions, such as the low noise Vicsek model, $\alpha$ and $\beta$ will change. In the specific case of low noise Vicsek data, points will accumulate along the $x=y$ and $x=y=w$ diagonals for $\miop$ and $\teop$, respectively. This is ideal as it results in minimal points in $A_N$---\ie, $\beta \approx 0$. Furthermore, the amount of points in $A_B$ will also be significantly reduced since only points near the corners---that is near $x \approx y \approx w \approx 0 = 2\pi$---will be within the required threshold, with $\alpha \approx \frac{5 \epsilon ^D}{N}$, noting that $\epsilon$ will not match \Eqnref{eq:avgDist} due to the distribution change.

Given that $k \ll N$, it is clear that \EqnTworef{eq:miComplexity}{eq:teComplexity} are of order $\mathcal{O}(N\tlogtwo N)$ and thus equivalent to the VP Tree. In practical terms, however, a non-zero $\alpha$ and $\beta$ as well as requiring twice the construction time---which is $\mathcal{O}(N\tlogtwo N)$ itself---will result in lower processing throughput.

To demonstrate the difference between the two approaches, empirical data is provided below where Vicsek data sets are analysed against the number of interactions generated. Only $\miop$ and $\teop$ are investigated as these have variable $N_I$, while $\gteop$ has constant $N_I=\tau M$ and thus does not provide any additional insight. As mentioned earlier, both algorithm implementations are provided with off the shelf software---with slight modification of the VP Tree implementation such that it uses contiguous memory allocation similar to the KD Tree implementation.

\begin{figure}[!h]
\centering
  \includegraphics[width=\columnwidth]{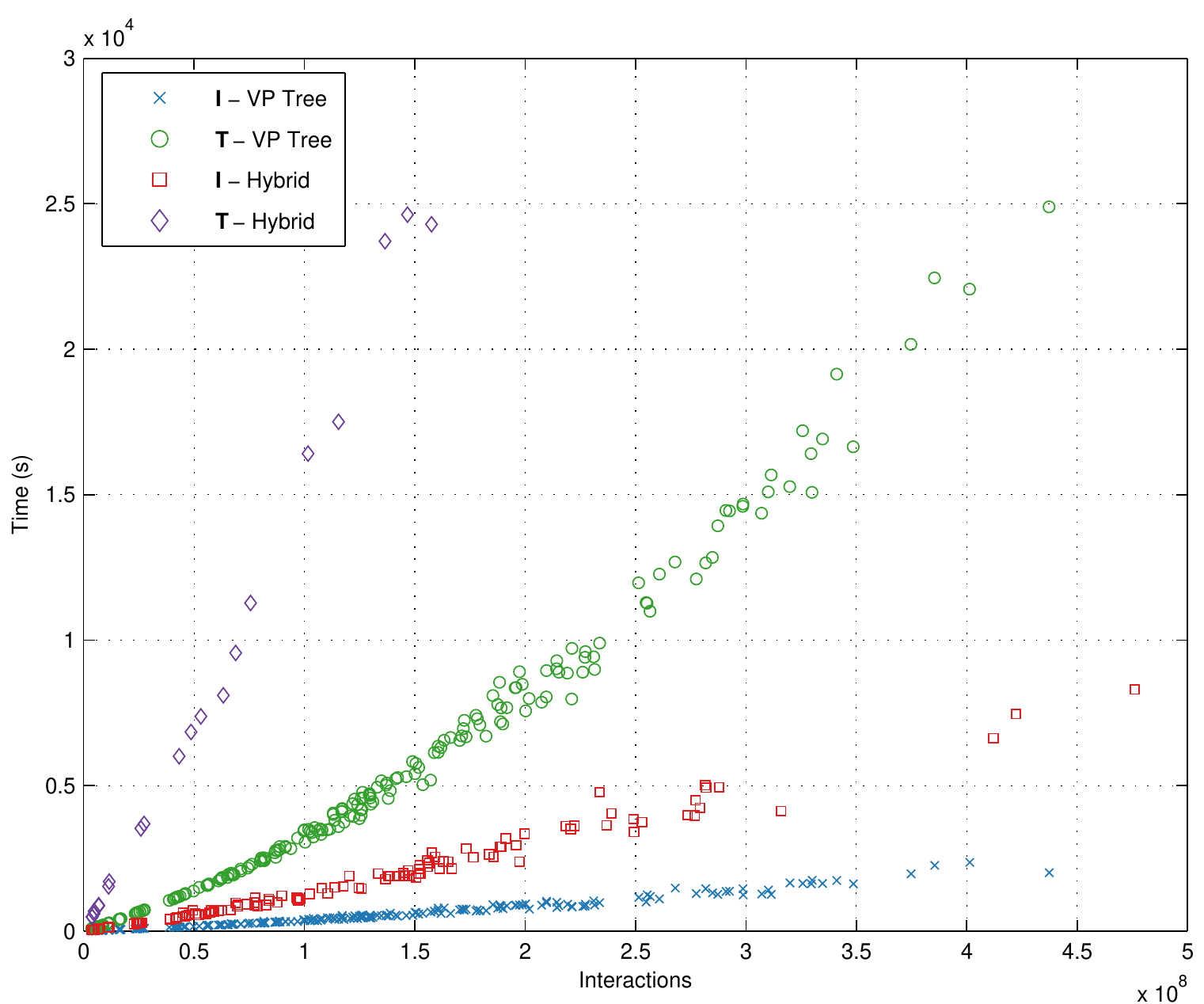}
  \caption{Time (s) vs $N_I$ measured for a Vicsek system with parameters: $N=1000, \tau=5000, \rho=0.25, s=0.1$ using VP Tree and hybrid methods.}
  \label{fig:timeUsage}
\end{figure}

\Figref{fig:timeUsage} shows the time taken to calculate $\miop$ and $\teop$ for a range of $N_I$. While all four measurements seem to scale according to $N_I \tlogtwo N_I$---as described above---the VP Tree methods are faster. The difference in running time for calculation of $\miop$ is strictly in finding the $\epsilon$-ball sizes, as both FR searches are performed with the same binary search function. Calculation of $\teop$ is worse still due to relying on the hybrid method more, with computations taking almost ten hours with the hybrid method compared to only 30 minutes under the VP Tree method.

\begin{figure}[!h]
\centering
  \includegraphics[width=\columnwidth]{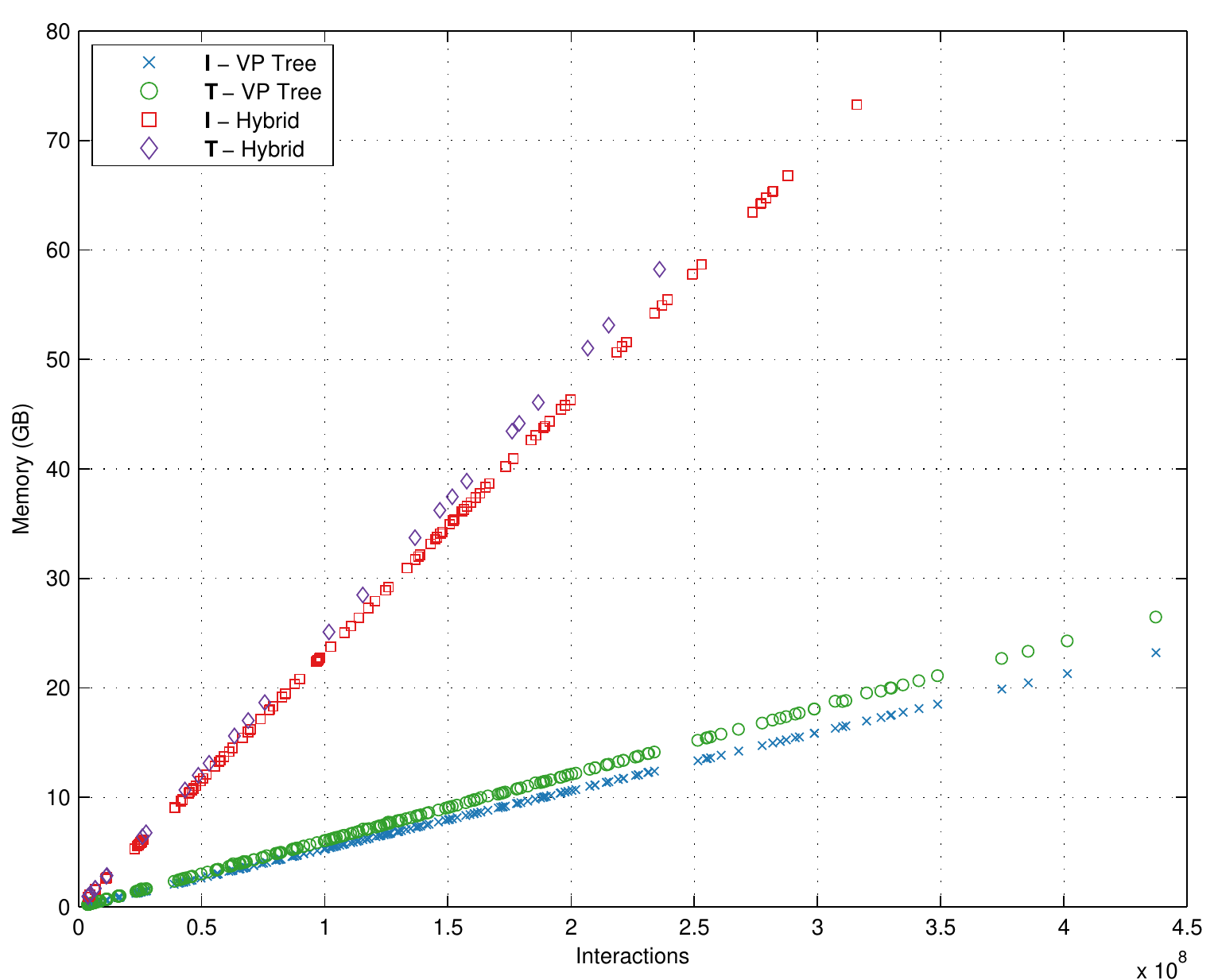}
  \caption{Memory usage (GB) vs $N_I$ measured for a Vicsek system with parameters: $N=1000, \tau=5000, \rho=0.25, s=0.1$ using VP Tree and hybrid methods.}
  \label{fig:memoryUsage}
\end{figure}

\Figref{fig:memoryUsage} shows the memory usage for the same calculations as above. All four metrics are rigidly linear in their scaling---as expected---but there is a marked increase in memory usage between the two methods, with the hybrid method performing definitively worse with both metrics.

\subsection{Additional checks of numerical stability}
\label{subsec:vicsekChecks}

Two additional checks for numerical stability were performed with the Vicsek data---a random shuffle and a data decimation. These additional checks were not performed for the canonical distributions as exact results are known for these.

\begin{figure}[!h]
\centering
  \includegraphics[width=\columnwidth]{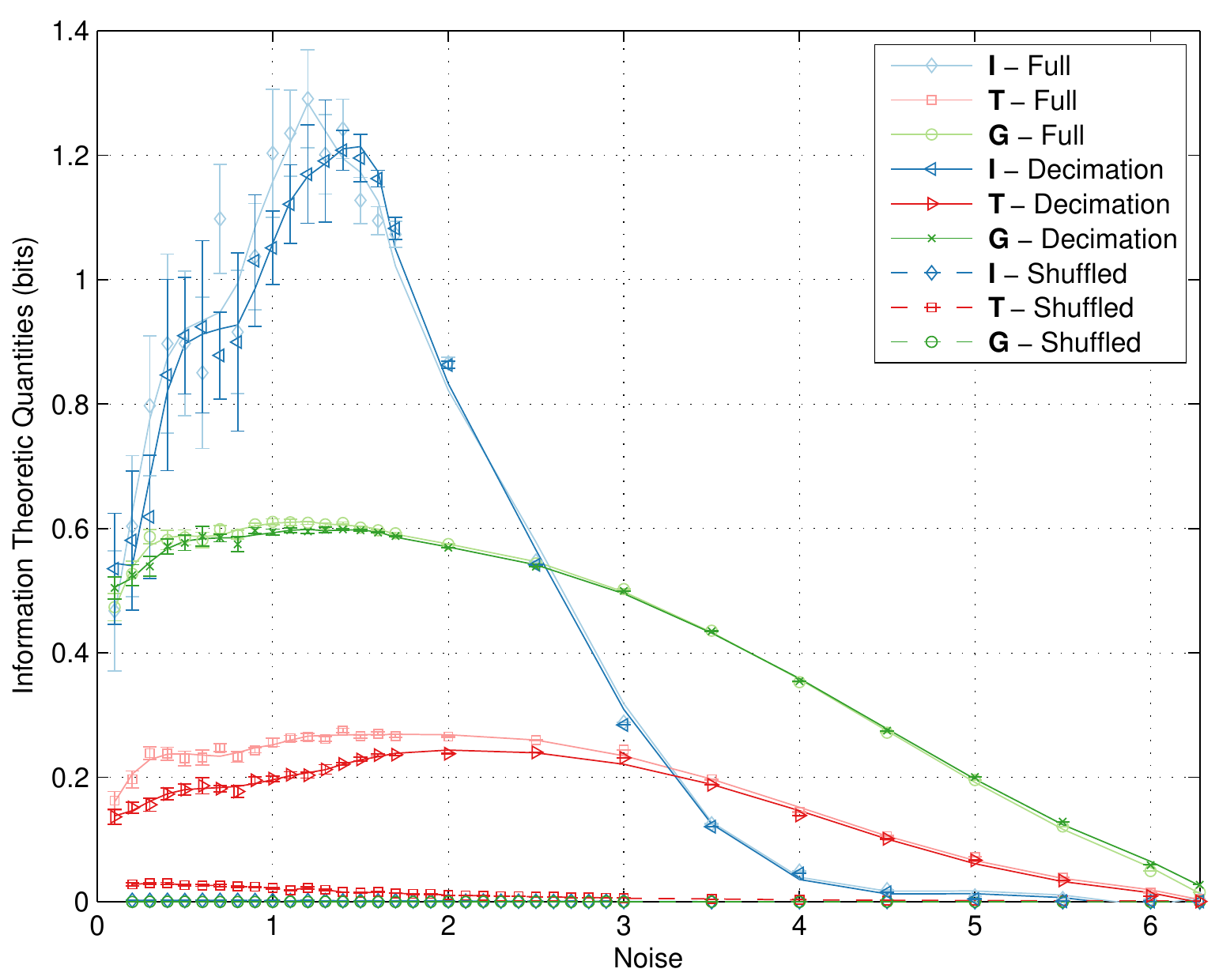}
  \caption{Results of random shuffle (dashed) and decimation (solid bold) for a Vicsek system with parameters: $N=1000, \tau=5000, \rho=0.25, s=0.1$. Results for unmodified data shown washed out. Shuffling has removed almost all information sharing between variables as evidenced by the near zero value for all three metrics over all noise. Decimation of data (Using random $10\%$ of available data) has little impact on results.}
  \label{fig:vicsek_shuffle_dec}
\end{figure}

In the first check, a random shuffle is employed on the coordinates corresponding to the $Y$ variable(s) for all three metrics. This should eliminate most of the information sharing between data sets and thus result in $\miop \approx \teop \approx \gteop \approx 0$. Results can be seen in \Figref{fig:vicsek_shuffle_dec}.

The decimation method is widespread and is employed to test the precision of the NN estimator implementation in a manner analogous to cross-validation~\citep{Witten05}. Specifically, data is generated from the Vicsek simulations, and a random subset containing 10\% of the entire population is chosen, which were then processed with the NN estimators. This was repeated ten times per run. Figure~\ref{fig:vicsek_shuffle_dec} shows that there was not a significant impact from this decimation, in alignment with assertions by \citet{Gao15}.

\subsection{Periodicity in data structure}
\label{subsec:periodicityImportance}

\citet{Hnizdo08} uses the \texttt{ANN} library for their KD Tree implementation, as here, in which they measure $\miop$ for the circular von Mises distribution. Interestingly however, they make no mention of the inability for KD Trees to handle periodic situations. Furthermore, their parameters for the von Mises distribution use a mean of $\pi$ for both random variables---with a periodic range of $[-\pi,\pi)$---meaning one would expect periodic artefacts to be quite pronounced as the distribution is centred on the border.

However, this is not the case. $500$ sets of $1\e4$ realisations of parameter set c (\Tabref{tab:ksgEstimateAndExact}) were tested with and without wrapping enabled in the VP Tree implementation. The estimated $\miop$ for the two tests were $0.194013$ and $0.194037$ for wrapped and non-wrapped VP trees, respectively, with an exact result of $0.1928$.

To see why the difference in results is not larger, consider not accounting for periodic conditions. This will only affect points near the boundary, where instead of counting the neighbours between $\pi \pm \epsilon(i)$---with wrapping handled---the process counts the neighbours in the range $[\pi-\epsilon(i)', \pi]$ (or $[-\pi, -\pi+\epsilon(i)']$), noting that $\epsilon(i)'$ will not necessarily be the same as $\epsilon(i)$ since it is likely the $k$th nearest  neighbour will be a different point. In a data set centred around $\pi$, even though many points are near the boundary, only a small subset will cross the boundary in either the kNN or FR searches. Additionally, the mirrored density on either side of the boundary means that $n_x'\approx n_x$ for these points---likewise for $n_y$. Note that the actual distance to the $k$th nearest neighbour is not used in \Eqnsref{eq:miKraskovI1}{eq:gteGomez}, only the neighbour counts to which the digamma function is applied---which is approximately logarithmic for values over 10. These facts, combined with the result being averaged over all $N$ leads to the small change in result when periodic boundaries are ignored.

To test for the mirrored density assertion above, the data set is shifted such that the data on the boundary is asymmetric. This is achieved using $\mu_1=\mu_2=\frac{7\pi}{8}$. The estimate for wrapped trees is (rightfully) unaffected, however the non-wrapped trees estimate worsens, to $0.194050$. 

\section{Conclusion}
\label{sec:conclusion}

In this paper, we discuss estimating information theoretic quantities for continuous variables using nearest neighbour estimators, specifically for systems with a large amount of circular data. The direct method of calculation is shown to be infeasible as it scales proportional to $N^2$. While spatial partitioning structures can reduce this to $N\tlogtwo N$, special care must be taken when dealing with periodic boundary conditions.

We discuss three data structures which are capable of handling these boundary conditions and have demonstrated that choosing a more appropriate data structure like the VP Tree can significantly increase performance in terms of time and memory used. Furthermore, as the VP Tree is fundamentally similar to the KD Tree, instead relying on distance between points rather than their specific topology, it should perform no worse than the KD Tree in any metric space on any manifold of a closed form, such as a sphere or \emph{Klein} bottle.

We have also shown that employing a decimation technique to reduce the number of points processed does not have a significant impact on estimation.

By combining the considerations put forth in this paper, estimation of $\miop$, $\teop$, and $\gteop$ for large systems with periodic boundary conditions is feasible and will show the same performance characteristics as systems with fewer constraints.

\section*{Acknowledgements}
Joshua Brown would like to acknowledge the support of his Ph.D. program and this work from the Australian Government Research Training Program Scholarship.

The National Computing Infrastructure (NCI) facility provided computing time for the simulations under project e004, with part funding under Australian Research Council Linkage Infrastructure grant LE140100002.

\vspace*{-3pt}   

\bibliographystyle{apa}

\begin{thebibliography}{33}
\expandafter\ifx\csname natexlab\endcsname\relax\def\natexlab#1{#1}\fi
\providecommand{\url}[1]{\texttt{#1}}
\providecommand{\href}[2]{#2}
\providecommand{\path}[1]{#1}
\providecommand{\DOIprefix}{doi:}
\providecommand{\ArXivprefix}{arXiv:}
\providecommand{\URLprefix}{URL: }
\providecommand{\Pubmedprefix}{pmid:}
\providecommand{\doi}[1]{\href{http://dx.doi.org/#1}{\path{#1}}}
\providecommand{\Pubmed}[1]{\href{pmid:#1}{\path{#1}}}
\providecommand{\bibinfo}[2]{#2}
\ifx\xfnm\relax \def\xfnm[#1]{\unskip,\space#1}\fi
\bibitem[{Barnett et~al.(2013)Barnett, Harr\'e, Lizier, Seth \&
  Bossomaier}]{Barnett13}
\bibinfo{author}{Barnett, L.}, \bibinfo{author}{Harr\'e, M.},
  \bibinfo{author}{Lizier, J.}, \bibinfo{author}{Seth, A.~K.}, \&
  \bibinfo{author}{Bossomaier, T.} (\bibinfo{year}{2013}).
\newblock \bibinfo{title}{Information flow in a kinetic ising model peaks in
  the disordered phase}.
\newblock {\it \bibinfo{journal}{Phys. Rev. Lett.}\/},  {\it
  \bibinfo{volume}{111}\/}, \bibinfo{pages}{177203}.
\bibitem[{Bentley(1975)}]{Bentley75}
\bibinfo{author}{Bentley, J.~L.} (\bibinfo{year}{1975}).
\newblock \bibinfo{title}{Multidimensional binary search trees used for
  associative searching}.
\newblock {\it \bibinfo{journal}{Communications of the ACM}\/},  {\it
  \bibinfo{volume}{18}\/}, \bibinfo{pages}{509--517}.
\bibitem[{Bhattacharyya \& Chakrabarti(2008)}]{Bhattacharyya08}
\bibinfo{author}{Bhattacharyya, P.}, \& \bibinfo{author}{Chakrabarti, B.~K.}
  (\bibinfo{year}{2008}).
\newblock \bibinfo{title}{The mean distance to the nth neighbour in a uniform
  distribution of random points: an application of probability theory}.
\newblock {\it \bibinfo{journal}{European Journal of Physics}\/},  {\it
  \bibinfo{volume}{29}\/}, \bibinfo{pages}{639}.
\bibitem[{Dwyer(1991)}]{Dwyer91}
\bibinfo{author}{Dwyer, R.~A.} (\bibinfo{year}{1991}).
\newblock \bibinfo{title}{Higher-dimensional voronoi diagrams in linear
  expected time}.
\newblock {\it \bibinfo{journal}{Discrete \& Computational Geometry}\/},  {\it
  \bibinfo{volume}{6}\/}, \bibinfo{pages}{343--367}.
\bibitem[{Friedman et~al.(1977)Friedman, Bentley \& Finkel}]{Friedman77}
\bibinfo{author}{Friedman, J.~H.}, \bibinfo{author}{Bentley, J.~L.}, \&
  \bibinfo{author}{Finkel, R.~A.} (\bibinfo{year}{1977}).
\newblock \bibinfo{title}{An algorithm for finding best matches in logarithmic
  expected time}.
\newblock {\it \bibinfo{journal}{ACM Transactions on Mathematical Software
  (TOMS)}\/},  {\it \bibinfo{volume}{3}\/}, \bibinfo{pages}{209--226}.
\bibitem[{Fuchs et~al.(1980)Fuchs, Kedem \& Naylor}]{Fuchs80}
\bibinfo{author}{Fuchs, H.}, \bibinfo{author}{Kedem, Z.~M.}, \&
  \bibinfo{author}{Naylor, B.~F.} (\bibinfo{year}{1980}).
\newblock \bibinfo{title}{On visible surface generation by a priori tree
  structures}.
\newblock In {\it \bibinfo{booktitle}{ACM Siggraph Computer Graphics}\/} (pp.
  \bibinfo{pages}{124--133}).
\newblock \bibinfo{organization}{ACM} volume~\bibinfo{volume}{14}.
\bibitem[{Gaboune et~al.(1993)Gaboune, Laporte \& Soumis}]{Gaboune93}
\bibinfo{author}{Gaboune, B.}, \bibinfo{author}{Laporte, G.}, \&
  \bibinfo{author}{Soumis, F.} (\bibinfo{year}{1993}).
\newblock \bibinfo{title}{Expected distances between two uniformly distributed
  random points in rectangles and rectangular parallelpipeds}.
\newblock {\it \bibinfo{journal}{Journal of the Operational Research
  Society}\/},  {\it \bibinfo{volume}{44}\/}, \bibinfo{pages}{513--519}.
\bibitem[{Gao et~al.(2015)Gao, Ver~Steeg \& Galstyan}]{Gao15}
\bibinfo{author}{Gao, S.}, \bibinfo{author}{Ver~Steeg, G.}, \&
  \bibinfo{author}{Galstyan, A.} (\bibinfo{year}{2015}).
\newblock \bibinfo{title}{Efficient estimation of mutual information for
  strongly dependent variables.}
\newblock In {\it \bibinfo{booktitle}{AISTATS}\/} (pp.
  \bibinfo{pages}{277--286}).
\bibitem[{G{\'o}mez-Herrero et~al.(2015)G{\'o}mez-Herrero, Wu, Rutanen,
  Soriano, Pipa \& Vicente}]{Gomez-Herrero15}
\bibinfo{author}{G{\'o}mez-Herrero, G.}, \bibinfo{author}{Wu, W.},
  \bibinfo{author}{Rutanen, K.}, \bibinfo{author}{Soriano, M.~C.},
  \bibinfo{author}{Pipa, G.}, \& \bibinfo{author}{Vicente, R.}
  (\bibinfo{year}{2015}).
\newblock \bibinfo{title}{Assessing coupling dynamics from an ensemble of time
  series}.
\newblock {\it \bibinfo{journal}{Entropy}\/},  {\it \bibinfo{volume}{17}\/},
  \bibinfo{pages}{1958--1970}.
\bibitem[{Hahs \& Pethel(2013)}]{Hahs13}
\bibinfo{author}{Hahs, D.~W.}, \& \bibinfo{author}{Pethel, S.~D.}
  (\bibinfo{year}{2013}).
\newblock \bibinfo{title}{Transfer entropy for coupled autoregressive
  processes}.
\newblock {\it \bibinfo{journal}{Entropy}\/},  {\it \bibinfo{volume}{15}\/},
  \bibinfo{pages}{767--788}. \DOIprefix\doi{10.3390/e15030767}.
\bibitem[{Hanov(2012)}]{Hanov12}
\bibinfo{author}{Hanov, S.} (\bibinfo{year}{2012}).
\newblock \bibinfo{title}{Vp trees: A data structure for finding stuff fast} \url{http://stevehanov.ca/blog/index.php?id=130}.
\bibitem[{Harr\'e \& Bossomaier(2009)}]{harre09:epl}
\bibinfo{author}{Harr\'e, M.}, \& \bibinfo{author}{Bossomaier, T.}
  (\bibinfo{year}{2009}).
\newblock \bibinfo{title}{Phase-transition -- behaviour of information measures
  in financial markets}.
\newblock {\it \bibinfo{journal}{Europhysics Letters}\/},  {\it
  \bibinfo{volume}{87}\/}, \bibinfo{pages}{18009}.
\bibitem[{Hertz(1909)}]{Hertz09}
\bibinfo{author}{Hertz, P.} (\bibinfo{year}{1909}).
\newblock \bibinfo{title}{{\"U}ber den gegenseitigen durchschnittlichen abstand
  von punkten, die mit bekannter mittlerer dichte im raume angeordnet sind}.
\newblock {\it \bibinfo{journal}{Mathematische Annalen}\/},  {\it
  \bibinfo{volume}{67}\/}, \bibinfo{pages}{387--398}.
\bibitem[{Hnizdo et~al.(2008)Hnizdo, Tan, Killian \& Gilson}]{Hnizdo08}
\bibinfo{author}{Hnizdo, V.}, \bibinfo{author}{Tan, J.},
  \bibinfo{author}{Killian, B.~J.}, \& \bibinfo{author}{Gilson, M.~K.}
  (\bibinfo{year}{2008}).
\newblock \bibinfo{title}{Efficient calculation of configurational entropy from
  molecular simulations by combining the mutual-information expansion and
  nearest-neighbor methods}.
\newblock {\it \bibinfo{journal}{Journal of computational chemistry}\/},  {\it
  \bibinfo{volume}{29}\/}, \bibinfo{pages}{1605--1614}.
\bibitem[{Ising(1925)}]{ising25}
\bibinfo{author}{Ising, E.} (\bibinfo{year}{1925}).
\newblock \bibinfo{title}{Beitrag zur theorie des ferromagnetismus}.
\newblock {\it \bibinfo{journal}{Z. Phys.}\/},  {\it \bibinfo{volume}{31}\/},
  \bibinfo{pages}{253--258}.
\bibitem[{Kozachenko \& Leonenko(1987)}]{Kozachenko87}
\bibinfo{author}{Kozachenko, L.}, \& \bibinfo{author}{Leonenko, N.~N.}
  (\bibinfo{year}{1987}).
\newblock \bibinfo{title}{Sample estimate of the entropy of a random vector}.
\newblock {\it \bibinfo{journal}{Problemy Peredachi Informatsii}\/},  {\it
  \bibinfo{volume}{23}\/}, \bibinfo{pages}{9--16}.
\bibitem[{Kraskov et~al.(2004)Kraskov, St\"{o}gbauer \&
  Grassberger}]{Kraskov04}
\bibinfo{author}{Kraskov, A.}, \bibinfo{author}{St\"{o}gbauer, H.}, \&
  \bibinfo{author}{Grassberger, P.} (\bibinfo{year}{2004}).
\newblock \bibinfo{title}{Estimating mutual information}.
\newblock {\it \bibinfo{journal}{Physical Review E}\/},  {\it
  \bibinfo{volume}{69}\/}, \bibinfo{pages}{066138--066153}.
  \DOIprefix\doi{10.1103/PhysRevE.69.066138}.
\bibitem[{Lee(1982)}]{Lee82}
\bibinfo{author}{Lee, D.-T.} (\bibinfo{year}{1982}).
\newblock \bibinfo{title}{On k-nearest neighbor voronoi diagrams in the plane}.
\newblock {\it \bibinfo{journal}{IEEE Transactions on Computers}\/},  {\it
  \bibinfo{volume}{100}\/}, \bibinfo{pages}{478--487}.
\bibitem[{Lizier et~al.(2012)Lizier, Atay \& Jost}]{liz12b}
\bibinfo{author}{Lizier, J.~T.}, \bibinfo{author}{Atay, F.~M.}, \&
  \bibinfo{author}{Jost, J.} (\bibinfo{year}{2012}).
\newblock \bibinfo{title}{Information storage, loop motifs, and clustered
  structure in complex networks}.
\newblock {\it \bibinfo{journal}{Physical Review E}\/},  {\it
  \bibinfo{volume}{86}\/}, \bibinfo{pages}{026110+}.
  \DOIprefix\doi{10.1103/physreve.86.026110}.
\bibitem[{Lizier et~al.(2008)Lizier, Prokopenko \& Zomaya}]{liz08a}
\bibinfo{author}{Lizier, J.~T.}, \bibinfo{author}{Prokopenko, M.}, \&
  \bibinfo{author}{Zomaya, A.~Y.} (\bibinfo{year}{2008}).
\newblock \bibinfo{title}{{Local information transfer as a spatiotemporal
  filter for complex systems}}.
\newblock {\it \bibinfo{journal}{Physical Review E}\/},  {\it
  \bibinfo{volume}{77}\/}, \bibinfo{pages}{026110+}.
\bibitem[{Matsuda et~al.(1996)Matsuda, Kudo, Nakamura, Yamakawa \&
  Murata}]{Matsuda96}
\bibinfo{author}{Matsuda, H.}, \bibinfo{author}{Kudo, K.},
  \bibinfo{author}{Nakamura, R.}, \bibinfo{author}{Yamakawa, O.}, \&
  \bibinfo{author}{Murata, T.} (\bibinfo{year}{1996}).
\newblock \bibinfo{title}{Mutual information of {I}sing systems}.
\newblock {\it \bibinfo{journal}{Int. J. Theor. Phys.}\/},  {\it
  \bibinfo{volume}{35}\/}, \bibinfo{pages}{839--845}.
\bibitem[{Mount \& Arya(2010)}]{Mount10}
\bibinfo{author}{Mount, D.~M.}, \& \bibinfo{author}{Arya, S.}
  (\bibinfo{year}{2010}).
\newblock \bibinfo{title}{Ann: A library for approximate nearest neighbor
  searching (2006)}.
\bibitem[{Ross(2014)}]{Ross14}
\bibinfo{author}{Ross, B.~C.} (\bibinfo{year}{2014}).
\newblock \bibinfo{title}{Mutual information between discrete and continuous
  data sets}.
\newblock {\it \bibinfo{journal}{PloS one}\/},  {\it \bibinfo{volume}{9}\/},
  \bibinfo{pages}{e87357}.
\bibitem[{Schreiber(2000)}]{Schreiber00}
\bibinfo{author}{Schreiber, T.} (\bibinfo{year}{2000}).
\newblock \bibinfo{title}{Measuring information transfer}.
\newblock {\it \bibinfo{journal}{Phys. Rev. Lett.}\/},  {\it
  \bibinfo{volume}{85}\/}, \bibinfo{pages}{461--464}.
\bibitem[{Shannon(1948)}]{Shannon48}
\bibinfo{author}{Shannon, C.} (\bibinfo{year}{1948}).
\newblock \bibinfo{title}{A mathematical theory of communication}.
\newblock {\it \bibinfo{journal}{Bell System Technical Journal}\/},  {\it
  \bibinfo{volume}{27}\/}, \bibinfo{pages}{379--423, 623--656}.
\bibitem[{Uhlmann(1991)}]{Uhlmann91}
\bibinfo{author}{Uhlmann, J.~K.} (\bibinfo{year}{1991}).
\newblock \bibinfo{title}{Satisfying general proximity / similarity queries
  with metric trees}.
\newblock {\it \bibinfo{journal}{Information Processing Letters}\/},  {\it
  \bibinfo{volume}{40}\/}, \bibinfo{pages}{175 -- 179}.
  \DOIprefix\doi{http://dx.doi.org/10.1016/0020-0190(91)90074-R}.
\bibitem[{Varilly(2014)}]{Varilly14}
\bibinfo{author}{Varilly, P.} (\bibinfo{year}{2014}).
\newblock \bibinfo{title}{Periodic kdtree}.
\newblock \URLprefix \url{https://github.com/patvarilly/periodic_kdtree}.
\bibitem[{Vicsek et~al.(1995)Vicsek, Czir\'{o}k, Ben-Jacob, Cohen \&
  Shochet}]{Vicsek95}
\bibinfo{author}{Vicsek, T.}, \bibinfo{author}{Czir\'{o}k, A.},
  \bibinfo{author}{Ben-Jacob, E.}, \bibinfo{author}{Cohen, I.}, \&
  \bibinfo{author}{Shochet, O.} (\bibinfo{year}{1995}).
\newblock \bibinfo{title}{Novel type of phase transition in a system of
  self-driven particles}.
\newblock {\it \bibinfo{journal}{Physical Review Letters}\/},  {\it
  \bibinfo{volume}{75}\/}, \bibinfo{pages}{1226--1229}.
  \DOIprefix\doi{10.1103/PhysRevLett.75.1226}.
\bibitem[{Wang et~al.(2012)Wang, Miller, Lizier, Prokopenko \& Rossi}]{wang12a}
\bibinfo{author}{Wang, X.~R.}, \bibinfo{author}{Miller, J.~M.},
  \bibinfo{author}{Lizier, J.~T.}, \bibinfo{author}{Prokopenko, M.}, \&
  \bibinfo{author}{Rossi, L.~F.} (\bibinfo{year}{2012}).
\newblock \bibinfo{title}{Quantifying and tracing information cascades in
  swarms}.
\newblock {\it \bibinfo{journal}{PLoS ONE}\/},  {\it \bibinfo{volume}{7}\/},
  \bibinfo{pages}{e40084+}. \DOIprefix\doi{10.1371/journal.pone.0040084}.
\bibitem[{Wicks et~al.(2007)Wicks, Chapman \& Dendy}]{Wicks07}
\bibinfo{author}{Wicks, R.~T.}, \bibinfo{author}{Chapman, S.~C.}, \&
  \bibinfo{author}{Dendy, R.} (\bibinfo{year}{2007}).
\newblock \bibinfo{title}{Mutual information as a tool for identifying phase
  transitions in dynamical complex systems with limited data}.
\newblock {\it \bibinfo{journal}{Physical Review E}\/},  {\it
  \bibinfo{volume}{75}\/}. \DOIprefix\doi{10.1103/PhysRevE.75.051125}.
\bibitem[{Witten \& Frank(2005)}]{Witten05}
\bibinfo{author}{Witten, I.~H.}, \& \bibinfo{author}{Frank, E.}
  (\bibinfo{year}{2005}).
\newblock {\it \bibinfo{title}{Data Mining: Practical machine learning tools
  and techniques}\/}.
\newblock \bibinfo{publisher}{Morgan Kaufmann}.
\bibitem[{Yianilos(1993)}]{Yianilos93}
\bibinfo{author}{Yianilos, P.~N.} (\bibinfo{year}{1993}).
\newblock \bibinfo{title}{Data structures and algorithms for nearest neighbor
  search in general metric spaces}.
\newblock In {\it \bibinfo{booktitle}{SODA}\/} (pp. \bibinfo{pages}{311--21}).
\newblock volume~\bibinfo{volume}{93}.
\bibitem[{Zhang et~al.(2013)Zhang, Huang, Geng \& Liu}]{Zhang13}
\bibinfo{author}{Zhang, Y.-m.}, \bibinfo{author}{Huang, K.},
  \bibinfo{author}{Geng, G.}, \& \bibinfo{author}{Liu, C.-l.}
  (\bibinfo{year}{2013}).
\newblock \bibinfo{title}{Fast knn graph construction with locality sensitive
  hashing}.
\newblock In {\it \bibinfo{booktitle}{Joint European Conference on Machine
  Learning and Knowledge Discovery in Databases}\/} (pp.
  \bibinfo{pages}{660--674}).
\newblock \bibinfo{organization}{Springer}.

\end{thebibliography}


\newcommand{\mybio}[2]{%
\begin{minipage}[t]{0.3\columnwidth}%
\vspace{0pt}%
\includegraphics[width=\columnwidth]{#1}%
\par \hfill%
\end{minipage} \hphantom{00}%
\begin{minipage}[t]{0.6\columnwidth}%
\vspace{0pt}%
{#2}%
\par \hfill%
\end{minipage}%
}

\mybio{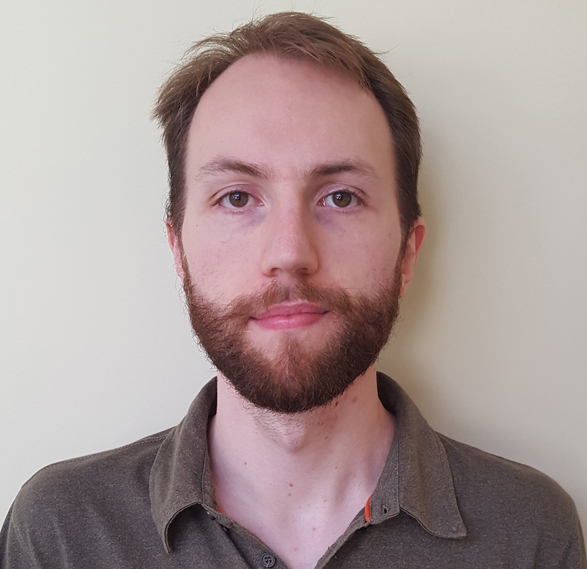}{{\bf Joshua Brown} received his Bachelor's degree (Hons) in Computer Science (Games Technology) in 2010 from Charles Sturt University, Bathurst, Australia. Currently he is a PhD candidate at Charles Sturt University, Bathurst, Australia. His research interests are computational modelling, optimisation, and information theory.}

\mybio{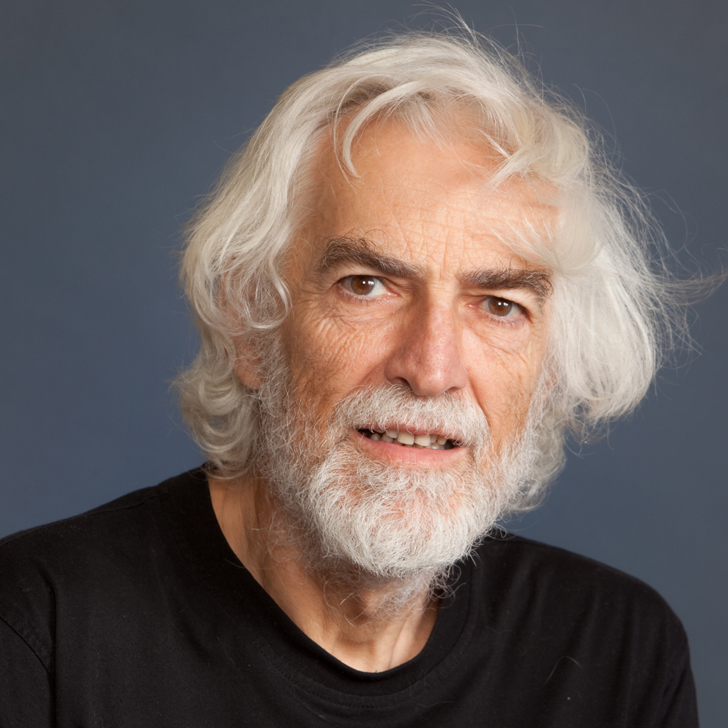}{{\bf Terry Bossomaier} is professor of computer systems at Charles Sturt University. His research interests range from complex systems to computer games. He has published numerous research articles in journals and conferences and several books, the most recent, on information theory appearing in December 2016.}

\mybio{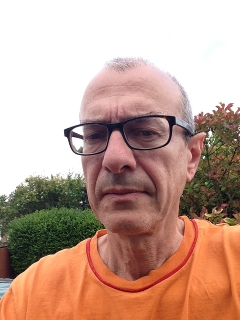}{{\bf Lionel Barnett} is a member of the Sackler Centre of Conciousness Science at the University of Sussex, United Kingdom. He has a number of publications with particular interest in evolutionary computation and Granger causality. He is also co-author with Terry Bossomaier of a Transfer Entropy book published in 2016.}

\end{document}